\documentstyle[aps,prb,preprint,psfig]{revtex}
\begin{document}
\draft

\title{Critical behaviour of thermopower and conductivity
at the metal-insulator transition in high-mobility Si-MOSFETs}

\author{R.\ Fletcher$^{a}$,  V.\ M.\ Pudalov$^b$,
A.\ D.\ B.\ Radcliffe$^a$ and C.\ Possanzini$^c$}
\address{$^a$ Physics Department, Queen's University, Kingston,
Ontario, Canada, K7L 3N6.}
\address{$^b$ P.\ N.\ Lebedev Physics
Institute, 117924 Moscow, Russia.}
\address{$^c$ Physics Department, University of Nijmegen,
625 ED Nijmegen, The Netherlands.}
\date{February 26, 2001}
\maketitle

\begin{abstract}

This paper reports thermopower and conductivity measurements
through the metal-insulator transition for 2-dimensional
electron gases in high mobility Si-MOSFETs.
At low temperatures both thermopower and conductivity
show critical behaviour as a function of electron density.
When approaching the critical density from the metallic
side the diffusion thermopower appears to diverge and the
conductivity vanishes. On the insulating side the thermopower
shows an upturn with decreasing temperature. These features
have much in common with those expected for an Anderson
transition.
\end{abstract}

\section{Introduction}
Scaling theory of non-interacting, disordered, electron gases
predicts that no metal-insulator transition (MIT) occurs in 2
dimensions\cite{scaling,scalingreview,interact_scaling} as
temperature $T \rightarrow 0$. Nevertheless, what appears to be
an MIT has been observed (at finite, though low $T$), first
in $n-$Si-MOSFETs\cite{MOSFETrefs} and more recently in many
other 2-dimensional (2D) hole and electron gases.\cite{otherMIT}
In the particular case of Si-MOSFETs, the transition is most
clearly visible in high-mobility samples, roughly $\mu \geq
1$\,m$^2$/V\,s. As the electron density, $n$, is varied, there is
a particular value, $n_0$, above or below which the resistivity
$\rho$ shows metallic or insulating temperature dependence
respectively. For the present purposes we will use as a working
definition that negative $d\rho/dT$ indicates an \lq insulator',
and positive $d \rho/dT$ at the lowest temperatures we can reach
corresponds to a \lq metal' (possible deviations from this
definition and the consequences will be mentioned later). At
$n>n_0$ and not too close to $n_0$, metallic behaviour is visible
over a wide range of $T$, roughly $T < 0.5 T_F$ where $T_F$ is the
Fermi temperature. The decrease of $\rho$ in the metallic state
for high mobility Si-MOSFETs samples is typically a few
orders of magnitude larger than can be accounted for by
electron-phonon scattering. There is no accepted explanation of
this behaviour as yet.  On the insulating side $\rho$ is found to
increase exponentially as $T$ is reduced, typically showing good
agreement with variable range hopping models.

Most previous work on these systems has focused on $\rho$,
though measurements of the compressibility\cite{compressibility}
have also appeared recently. The present paper presents
experimental data on thermopower, $S$, and
conductivity, $\sigma =1/\rho$, at temperatures down to 0.3\,K.
When we extract the diffusion thermopower, $S^d$, we find that
it diverges when approaching $n_0$. At $n_0$ there is an abrupt
change in behaviour of the thermopower with lower densities showing
an upturn in $S$ as $T$ is decreased. We also attribute this to $S^d$.

In addition we find that $\sigma$ exhibits critical behaviour
around $n_0$ which appears to be largely consistent with the
behaviour of $S^d$. Earlier, a scaling behaviour was
described\cite{MOSFETrefs} for the temperature dependence of
$\rho(T)$ over a temperature range $\sim (0.05 - 0.3) T_F$. In
contrast, the present experiments are concerned with the limiting,
low temperature behaviour of $\sigma$. As already mentioned,
$\rho$ decreases with temperature when $n > n_0$ but it saturates
at low temperatures. (We are unable to say what happens below
0.3\,K and return to this point later). Our low temperature
saturation values of $\sigma$, say $\sigma_0$, show a power-law
critical behaviour as a function of $(n/n_0-1)$. Similar critical
behaviour of $\sigma$ around $n_0$ has previously been  reported
in 2D p-type GaAs\cite{Hanein} and lower mobility
Si-MOSFETs.\cite{Feng} The observed characteristics of $S^d$ and
$\sigma_0$ are strongly reminiscent of those expected for an
Anderson MIT in 3D\cite{Mott,Castellani,Enderby,Villagonzalo} but
such a transition should not occur in 2D.

We have reported some of
our preliminary results, particularly on the thermopower, in
conference form.\cite{EP2DS} The present paper extends the
thermopwer results to a wider range of temperatures and densities,
and extends the data into regions of high sample resistivities
which were inaccessible to us earlier. It is interesting to note
that, unlike resistivity, $S$ is measured with no current flow
in the system and hence no self heating occur.
We also present the results
of systematic studies on $\rho$ on a number of samples; these were
primarily taken to augment our understanding of the thermopower,
but they are also interesting in their own right.

\section{Sample and Experiments}
The main sample used for the present $\rho$ and $S$ measurements
(Sample 1) is the same as that described in a previous
paper\cite{paper1} and the general techniques used to  measure $S$
can also be found there. This sample has $n_0 = 1.01 \times
10^{15}$\,m$^{-2}$ (as defined as above) and a peak mobility
$\mu = 1.82$\,m$^2$/Vs at $T = 0.3$\,K. $S$ and $\rho$ have been
measured as a function of $T$, down to about
0.3\,K, at many different values of $n$. We have also analyzed
independent $\rho(T,n)$ data for two other samples over the same
range of $T$, Sample 2 from the same wafer with $n_0 = 0.96 \times
10^{15}$\,m$^{-2}$  and with peak $\mu = 1.96$\,m$^2$/Vs (again at
$T=0.3$\,K) and Sample 3 \cite{Pudalov}
with $n_0 = 0.96 \times 10^{15}$\,m$^{-2}$ and
peak $\mu = 3.6$\,m$^2$/Vs.
All connections to the 2D gas had isolation resistance $
>50\,$G$\Omega$ and all leads into the cryostat were well shielded
and filtered against rf interference. For the sample leads,
commercial, bulkhead, rf rejection filters were used covering a
range 1\,MHz - 10\,GHz, and these were augmented by extra, series
inductors to extend the range down to about 50\,kHz.

Measurements were made at dc. With the sample in the metallic
state ($n>n_0$), a Keithley 182 digital voltmeter usually gave the
best compromise of input bias current, input impedance and noise.
As previously,\cite{paper1} we obtained spurious voltages from the
sample even with no temperature gradient or current present. We
traced this to the input bias current of the voltmeter. With any
amplifier, the input bias current combined with the source
impedance produces spurious (offset) voltages which must be
separated from the true signals. Spurious voltages have been
noticed previously by many people and probably have a number of
causes, but we verified the consistency of the explanation in the
present case by measuring the 2-terminal resistance of the sample
over a wide range of conditions, as well as measuring the input
bias current of the amplifier. The bias current of the input
amplifier of our  Keithley 182 was $15$\,pA for sources up to
about 1\,M$\Omega$ but it increased to $30$\,pA at higher source
impedances.

When the system was metallic the offset voltages were always
rather small and not usually too $T$ dependent, especially at
lower temperatures where the thermoelectric signals are also
small, so they were not a serious problem. For example, at $ n =
2.1\times 10^{15}$m$^{-2}$, the sample is metallic and has a
resistance of $\sim 20$\,k$\Omega$ at low temperatures. This gives
an offset voltage of $\sim 30$\,nV. Offsets were taken into
account by taking the thermoelectric signal to be the difference
between the measured voltages with the temperature gradient
present and not present, ensuring that the mean sample
temperature, and thus resistance, was the same in both cases.

On the other hand, when the sample was in the insulating state its
resistance was much higher and strongly $T-$dependent. In this
case the offset voltage rapidly overwhelmed the thermoelectric
signal as the temperature was lowered. To reduce this problem, an
amplifier with input bias current $< 1$\,pA and input impedance
$>10^{12}\Omega$ was used. With some averaging it had a resolution
of 0.1\,$\mu$V for source impedances of less than a few hundred
k$\Omega$ rising to about 1\,$\mu$V at 10-20\,M$\Omega$. However,
when used with the sample there was always an apparent offset
current of $\sim4$\,pA independent of sample voltage, i.e., offset
signals were produced corresponding to a dc current through the
sample of this magnitude. We were unable to determine the origin
of this current. It was unaffected by changing the rf filters in
the leads, or by rerouting some of the wiring, particularly that
carrying the gate voltage. When the offset signals become large,
one would like to have some check that the method of extracting
the signal used above was reliable. Fortunately, it was found that
if the sample was grounded at different contacts, the offset
current always flowed in a direction towards ground. Thus by
measuring the thermoelectric signals with the source or drain
grounded in turn, the offset voltage reversed sign but the
thermoelectric voltage did not. This gave two independent
measurements and is the origin of the error bars on our data at
low temperatures with the sample in the insulating state (see
Fig.~(4) later). The width of the error bars typically increases
very rapidly as $T$ is lowered because of the strong variation of
$\rho$.

Finally we provide some information on the temperature difference
$\Delta T$ across the sample at various temperatures so that the
signal voltages $\Delta V = S \Delta T$ may be estimated under various
conditions.
The electric field in the sample may also be estimated given the length
was 2.5\,mm. Between 0.7 and 7\,K, $\Delta T \approx 100$\,mK. Below
0.7\,K $\Delta T$ dropped steadily until by 0.4\,K it was about 30\,mK
and at 0.34\,K about 15\,mK.

\section{Results and Discussion}
In the metallic region $n$, as obtained from the Hall constant,
was found to be a linear function of gate voltage $V_g$, and it is
believed to follow approximately the same dependence in the
insulating region,\cite{hall} at least close to $n_0$. On
different cooldowns from room temperature there were small shifts
in $n$ at fixed $V_g$ of up to $0.04\times 10^{15}$\,m$^{-2}$, but
if the sample was held at 77\,K between cooldowns, as was done for
most of these measurements, the density and other properties were
very reproducible.
In general the absolute uncertainty in $n$ should be no worse than
$\pm 0.1\times 10^{15}$\,m$^{-2}$ in the region of $n_0$ and 2\%
at high $n$; the reason for the larger uncertainty near $n_0$ is
the fact that the Hall resistivity shows non-linearities of up to
about 10\% there.

Because our analysis of thermopower makes extensive use of $\rho$
data, examples of the temperature dependence of $\rho$ on both the
metallic and insulating side are shown in Figs.~1 and 2.  They are
very similar to those seen in
previous work.\cite{MOSFETrefs,VRHresistivity}

Well into the insulating regime (roughly $n \leq 0.65\times
10^{15}$m$^{-2}$) $\rho$ can be well represented by an equation
developed for variable range hopping (VRH) through localized states
in the presence of a Coulomb gap at the Fermi energy $E_F$
(referred to as Efros-Shklovskii VRH),
\begin{equation}
\label{EFVRH} \rho = \rho_c \exp {(T_{ES}/T)^{1/2}},
\end{equation}
where $\rho_c$ and $T_{ES}$ are constants that depend on $n$ but
not $T$. Examples are given in Fig.~1 for Sample 1. The values
that we obtain for the constants are similar with those determined
by others.\cite{VRHresistivity} $T_{ES}$ varies from $\sim 35$\,K
at $n = 0.650\times 10^{15}$m$^{-2}$ to $\sim 90$\,K at
$0.48\times 10^{15}$m$^{-2}$. If the exponent is left as a
variable we find an average value of $0.50 \pm 0.05$ over this
same range of $n$.  We find poorer fits assuming Mott VRH
(no Coulomb gap at $E_F$) which corresponds to a power law of 1/3
in the above expression.

For  $n \geq 0.70\times 10^{15}$m$^{-2}$, a simple activated
behaviour
\begin{equation}
\label{activated}
 \rho = \rho_c \exp {(T_A/T)}
\end{equation}
provides a very good fit to the data below about 1\,K. Thus these
results suggest a transition from simple activated behaviour to VRH
as $n$ decreases.

In the metallic region ($n>n_0$) we have fitted our data
to the equation\cite{Pudalov}
\begin{equation}
\label{metalrho}
  \rho = \rho_0 + \rho_1 \exp \left[-(T_0/T)^q\right]
\end{equation}
where $\rho_0$, $\rho_1$, $T_0$ and $q$ are constants that depend
on $n$ but not $T$.  Examples are shown in Fig.~2 for Sample 2.
For $n \geq 1.5\times 10^{15}$m$^{-2}$, excellent fits were
obtained by fixing $q=0.5$. For lower $n$, the values of $q$ (if
this is left as a variable) increased and approached 2.0 at $n
\approx 1.0\times 10^{15}$m$^{-2}$. The model curves fit the data
successfully over the range 0.3\,K (our lowest $T$) to about
$0.3\,T_F$. ($T_F = 7.3n$\,K with $n$ in $10^{15}$\,m$^{-2}$
units). Above this region the curves usually exhibit a
maximum.\cite{akk,amp2000} $T_0$ also provides a natural scale for
the data and, since $T_0$ decreases as  $n_0$ is approached
(Fig.~2 indicates $T_0$ for the lowest densities), a lowering of
the upper temperature limit is also consistent with this. The
reduced range of fitting at low $n$ means that, $\rho_1$, $T_0$
and $q$ are not well known here but, even so, $\rho_0$ retains a
relatively low error.

Figure~3 shows the results on $\sigma_0 = 1/\rho_0$ as a function
of $n$.\cite{rho_0_definition}
For all three samples $\sigma_0$ follow the critical behaviour
\begin{equation}
\label{nscalesigma}
\sigma_0 = \sigma_m + \sigma_s\left(\frac{n}{n_0} - 1 \right)^\nu.
\end{equation}
The solid lines are the best fits with the following parameters,
with $\sigma$ in units of $e^2/h$ (the data points at the nominal
values of $n_0$ are $\sigma$(0.3K) and were not used in the fitting).
For sample 2, $\sigma_m = 0.0\pm 0.3$, $\sigma_s = 13.6\pm 0.7$
and $\nu=0.94\pm 0.03$. A good fit was not possible over the whole
range of $n$ and so was restricted to $n < 3.0\times
10^{15}$\,m$^{-2}$ (a range of $(n/n_0 -1)$ of about 0.03 to 3).
Sample 1 has an identical behaviour within experimental error. The
higher-mobility Sample 3 also follows the same equation with
$\sigma_m=0.36\pm 0.15$, $\sigma_s = 34\pm 5$ and $\nu=1.39\pm
0.05$. We have a smaller range of $n$ for this sample, but the fit
still covers a range $(n/n_0 -1) \approx 0.02-0.4$. These results
suggest $\nu $ increases with peak mobility but clearly more data
on a variety of samples are required. The values of $\sigma_m$ for
Sample 1 and 2 are consistent with zero within experimental
uncertainty. The fits are good over the whole range except
for the last points near $n = n_0$ which do not fall on the curves.
This is not necessarily unexpected because a value of
$\sigma =0$ is not possible at a finite temperature due to thermal
broadening of the electronic distribution. Small errors in $n_0$
will have a similar effect.
For Sample 3, $\sigma_m$ may be finite. However, if
$n_0$  is allowed to decrease from  $0.96$ to about $0.93 \times
10^{15}$\,m$^{-2}$, a fit which is essentially indistinguishable
over the range of the data can also be obtained with
$\sigma_m =0.0 \pm 0.2$, $\sigma_s = 32\pm 5$ and $\nu=1.48\pm 0.05$;
both fits are actually shown on Fig.~3.
A small
discrepancy in $n_0$ could easily arise from the identification of
the critical density for the MIT with that density, $n_0$, where
$d\rho/dT$ changes sign, a procedure which has no firm physical
foundation;\cite{akk} in addition there are uncertainties and
possibly non-linearities in the determination of $n$ from gate voltage
values as mentioned above.\cite{hall}

 The critical behaviour described by
Eq.~(\ref{nscalesigma}) with $\sigma_m = 0$ is formally the same
as that expected for a (continuous) Anderson transition with a \lq
mobility edge' at $n_0$, whereas a finite $\sigma_m$ would
correspond to a (discontinuous) Mott-Anderson transition; neither
transition should arise in a non-interacting 2D
gas.\cite{scaling,scalingreview} Our $\rho$ data in the insulating
region are also consistent with this scenario in the sense that we
observe activated behaviour for $n$ below $n_0$. Similar
critical behaviour, usually with $\sigma_m$ consistent with zero,
has been seen in many 3D systems, typically with
values\cite{Villagonzalo} of $\nu$ in the range $0.5-1.3$. There
are also two previously reported cases related to 2D. Hanein {\em
et al.}\cite{Hanein} have made a similar analysis to the one above
for a 2D hole gas in GaAs and found a linear relation between
$\sigma_0$ and $n$, but with a finite $\sigma_m$. Feng {\it et
al.}\cite{Feng} also found critical behaviour in $\sigma_0$ in a
low mobility Si-MOSFET though we note that their data all show
negative $d\rho(T)/dT$, even in the range which they identify with
\lq metallic' behaviour. They found $\sigma(T)$ {\it increased}
with $T$ and fitted a $T^2$ dependence. This behaviour was
attributed to local momemts and is so different from that found
here that the two cases may not be closely related. Recently
\cite{Meir} a
model based on percolation of non-interacting electrons through
local quantum point contacts also led to zero-temperature
conductivity consistent with our observations.

Note that the data presented in Fig.~3 were obtained from
extrapolating $\sigma$ to $T=0$ by a procedure which focusses only
on the \lq strong' exponential, $T-$dependence of Eq.~(3). It
ignores any other weaker dependences\cite{rho_0_definition} that
may be present below $T=0.3$K, including those due to weak
localization and screening.

We now turn to the thermopower data. A selection of data on $S$ as
a function of $T$ is shown in Fig.~4. From our previous
work\cite{paper1} we know that the diffusion thermopower, $S^d$,
is almost zero at $n = 8.5\times 10^{15}$\,m$^{-2}$ and one sees
only phonon drag, $S^g$ which varies approximately as $T^6$ at
the lowest temperatures. The fact that $S^d$ is very low
at high $n$ was predicted by Karavolas and
Butcher\cite{KB91} who attributed it to a particular combination
of scattering mechanisms. Indeed, when $n \gg n_0$ the behaviour
of both $S^d$ and $S^g$ appears to be in generally good agreement
with theory.\cite{paper1}

The $T^6$ dependence of $S^g$ obtains only in the Bloch limit, i.e.,
when $q \ll 2k_F$ where $q$ is the average phonon wave number and $k_F$
the Fermi wave number. At $n = 8.5\times 10^{15}$\,m$^{-2}$ the condition
$q \ll 2k_F$
is satisfied below about 1.4\,K. Above this temperature one sees a
gradual decrease in the exponent to roughly $S^g \propto T^3$ by 4\,K,
but no simple power law is expected in this region.

As $n$ decreases, $S$ begins to show two distinct regions with different
$T$ dependences. At $T < 1$\,K, $S$ has a much weaker, approximately
linear, $T$-dependence indicative of $S^d$ becoming dominant. For
$n < n_0$, this low-$T$ dependence, which is characteristic of ordinary
metals, is replaced by an upturn in $S$; we will return to this feature later.

Concentrating first on the metallic region, we analysed the data
over the range 0.3-1.5\,K to obtain $S^d$ by assuming
\begin{equation}
\label{Sd+Sg}
  S = S^d + S^g = \alpha T + \beta T^s
\end{equation}
where $\alpha \ (= S^d/T$) and $\beta$ are constants that depend
on $n$. For the 2 highest densities we obtained good straight
lines by plotting $S/T$ as a function of $T^{s-1}$ with $s=6$ as
expected. At lower $n$, proportionately lower temperatures are
required to reach the Bloch limit and so we no longer expect to
see $S^g \propto T^6$ in our temperature range. Hence we simply
used $s$ as a variable to give the best fit.  For $n = 1.9-5
\times 10^{15}$\,m$^{-2}$ we found $s=4$, and for $ n =
0.97-1.5\times 10^{15}$\,m$^{-2}$, $s=5$. That $s$ seems to
increase again at the lowest $n$ is unexpected, but more data are
required before this is taken to be an experimental fact. We have
no reason to believe that $S^g$ is not well-behaved in this
temperature, density and resistivity range, but we must
caution that our understanding of $S^g$ in these circumstances is
not well founded as yet.\cite{EP2DS,ql} Given this it is important
to note that the value of $\alpha$ depends only weakly on the
choice of $s$. Indeed, providing $S^g$ has a much stronger $T$
dependence than linear, then simply taking the measured values of
$S/T$ at 0.3\,K to be equal to $\alpha$ gives values in good
agreement with those obtained using Eq.~(\ref{Sd+Sg}).

Fig.~5 shows that $\alpha$ as a function of $n$ appears
to diverge as $n \rightarrow n_0$. One would expect such a divergence
if $E_F$ approaches a band edge or a gap in the DOS giving
$S^d \propto 1/n$. Das Sarma {\it et al.}\cite{DasSarma} have
suggested that a similar mechanism, carrier freezeout, is responsible
for the MIT in 2DEGs. However, the present results are inconsistent
with these explanations because Hall data, both our own and those in
Ref.\onlinecite{hall}, show that in the vicinity of $n_0$
the mobile carrier density is not approaching zero but
equals $n$ within a few percent; also any temperature dependence of
$n$ at fixed $V_g$ is very small in the range $0.3-4.2$\,K.
Given that theory successfully
predicted\cite{KB91} $S^d$ to be zero at high $n$ as mentioned
above, it is possible that the present results have an explanation in
terms of the specific scattering processes important at low $n$, but
there is no theory available for this situation at present.

However, Eq.~(\ref{nscalesigma}) also implies a divergence of $S^d$.
With the assumption of a constant density of states (DOS),
Eq.~(\ref{nscalesigma}) is consistent with
\begin{equation}
\label{escalesigma}
\sigma(E_F) = \sigma_m + \sigma_s \left(\frac{E_F}{E_c} - 1 \right)^\nu .
\end{equation}
Again, with $\sigma_m =0$ this
is formally equivalent to an Anderson transition with $E_c$ being
the mobility edge. Taking $\sigma_m =0$, the use of the Mott relation
$S^d = -(\pi^2 k_B^2 T/3e)(\partial \ln \sigma/\partial E)_{E_F}$
with Eq.~(\ref{escalesigma}) then gives \cite{Castellani,Villagonzalo}
\begin{equation}
\label{metalsd}
 S^d = -\frac{\nu \pi^2 k_B^2 T}{3e(E_F- E_c)}~.
\end{equation}

The use of the Mott relation, and hence this result, is valid only
if $(E_F - E_c)/k_BT \gg 1$; in the opposite limit $S^d$ tends to
a constant\cite{Enderby,Villagonzalo} ($\sim 228\,\mu$V/K in 3D).
The saturation is a direct consequence of the fact
that the contribution to the conductivity of electrons below $E_c$
is zero in this model. The relevant
integral\cite{Castellani,Villagonzalo}  for $S^d$ has a weighting
factor $\sigma(E)(E-E_F)$ which, under normal conditions, leads to
$S^d$ being a measure of the derivative of $\sigma(E)$ with
respect to energy, and hence to the Mott relation. This is
consistent with Eq.~(\ref{metalsd}). However, when $E_F = E_c$
only electrons above $E_c$ contribute to the integral, which leads
to a constant value of $S^d$. Numerical
calculations\cite{Villagonzalo} show that the approximation of
Eq.~(\ref{metalsd}) gives a magnitude roughly a factor of 2 too
large when $(E_F - E_c)/k_BT \approx 2$ (i.e. because of the
approach to saturation). For our sample at $T = 0.4$\,K this
corresponds to $(n-n_0)/n_0 = \Delta$ (say) $ \approx 0.11$, using
the ideal DOS, $g_0$, with an effective mass of $0.19\,m_0$. If
the present  model was indeed appropriate, it suggests that we
should see saturation of $S^d$ as $T$ increases for the samples
with $n = 0.97$ and $1.06 \times 10^{15}$\,m$^{-2}$ (for higher
$n$ any saturation would be masked by the rapidly increasing
contribution from $S^g$). We are unable to see any obvious
evidence for saturation at these densities and  Eq.~(\ref{Sd+Sg})
still seems to be valid there. However, we should recall that
$S^g$ is not yet understood\cite{ql}  at low $n$ and high $\rho$
and there remains the possibility that an unexpected behaviour of
$S^g$ could mask important features in $S^d$.

Proceeding with the model, we have fitted Eq.~(\ref{metalsd})
to the measured $\alpha$, allowing for the possibility of
saturation as $n_0$ is approached.
To simulate a saturation we add $\Delta$ in the denominator
(but allow it to be a variable when determining the best
fit to the data) and, rewriting Eq.~(\ref{metalsd}) in terms
of $n$, we have
 \begin{equation}
 \alpha = S^d/T  =  -K /\sqrt{\Delta^2  + (\frac{n}{n_0} - 1)^2 }
\label{scalesd}
\end{equation}
where $K = \nu \pi^2 k_B^2 /(3eE_c)$ is a constant expected
to be about 32\,$\mu$V/K$^2$ for Sample 1, again using $g_0$.
If $\sigma_m$ is finite in Eq.~(\ref{escalesigma}), then the Mott
relation suggests that it will contribute to the denominator of
Eq.~(\ref{metalsd}), also softening the divergence at $n=n_0$. However,
the experimental $\sigma_m$ is so small that this is probably negligible
compared to the finite-$T$ effect considered here.

The best fit of the data to Eq.~(\ref{scalesd}) is shown by the
solid line in Fig.~5 and uses $\Delta = 0.15\pm 0.01$ and
$K = (9.5\pm 1.5)\,\mu$V/K$^2$. Only data with
 $n < 4 \times 10^{15}$\,m$^{-2}$ are fitted, a range
 consistent with that used for $\sigma_0$.
 $\Delta$ is consistent with that expected from the argument above,
but $K$ is too small by a factor of about 3.  As with
$\sigma$, the fit can be improved if $n_0$ is decreased, perhaps
by $0.05 \times 10^{15}$\,m$^{-2}$, in which case $\Delta$
decreases and may even become zero.

In spite of the relatively good correspondence between the model
and the data, we emphasize that we should be cautious in
necessarily concluding that the model is fundamentally correct. We
are comparing our results on a 2D system with a theoretical model
of an Anderson MIT believed to be appropriate only to
non-interacting electrons in 3D. At this point it is difficult to
know  whether the situation in 2D might be significantly changed
by including interactions along with the strong disorder; this is
a complex and ongoing theoretical problem (e.g. see Refs.
\onlinecite{scalingreview,interact_scaling} and references
therein). Some progress has been made on calculating $S^d$ with
the inclusion of weak interactions and disorder.\cite{reizer} We
are, however, unable to explain the observed strong $n$ dependence
in terms of the calculations which predict universal values.
The calculated corrections
are logarithmic in $T$. These would be the equivalent of similar
corrections in $\rho$ and, if they exist, would require much lower
temperatures than we have available and would be difficult to
detect. On the other hand, these facts mean that the analysis of
$\alpha $, $\rho_0$, and their relationship as zero-temperature
quantities, is a self-consistent procedure which has a
semi-classical meaning.  Nevertheless, whether
or not $\sigma$ goes sharply to zero near $n_0$ as would be
required in the Anderson model, the observed change in behaviour
of $S^d$ as $n$ is varied around $n_0$ points to a fundamental
change in transport taking place in this region.

We should finally mention that we can also represent the data
reasonably  well over the same range using the simple expression
$\alpha = -56/n^{2.5}\,\mu$V/K$^2$, with $n$ in units of
$10^{15}$\,m$^{-2}$, but this has no obvious physical explanation;
in particular, it does not
have the form that we would expect for $S^d$ approaching a band edge at
$n = 0$, i.e. $\alpha \propto 1/n$.

Our data in the insulating regime also show a behaviour qualitatively
consistent with activation across an energy or mobility gap.
In such cases we expect an upturn of $S^d$ at low temperatures according
to
\begin{equation}
\label{insulsd}
S^d = -\frac{k_B}{e}\left(A + \frac{E_c - E_F}{k_BT} \right)
\end{equation}
where $A$ is a constant of order unity. In the particular case of
the Anderson model\cite{Mott,Villagonzalo,I&S} $Ak_B/e$
corresponds to the saturation value when $E_c = E_F$ noted above
in our discussion of the metallic region. When $n$ is close to, but
less than, $n_0$, we see activated behaviour in $\rho$
(see Eq.~(\ref{activated}))
which is in accord with this.
We do not see a high temperature saturation of $S$ as
Eq.~(\ref{insulsd}) predicts, but this is not surprisingly
because it will be masked by other contributions to $S$,
particularly $S^g$ (see also below).

There might also be another contribution to $S^d$ from VRH.
Demishev {\it et al.}\cite{Demishev} have recently demonstrated
the existence of such a component in $S^d$ using a 3D GaSb sample.
The $T$ dependence of our $\rho$ data and other previously
published data\cite{VRHresistivity} well into the insulating
region are consistent with Efros-Shklovskii VRH. (When two or more
conduction mechanisms are present, the appropriate $S^d$ are
weighted by their contributions to $\sigma$). For this mechanism
one expects $S^d$ to be a constant given by\cite{B&CVRH,Zvyagin}
$S^d = -(k_B/e)(k_BT_{ES}/C)(\partial \ln g(E)/\partial E)_{E_F}$
where $T_{ES}$ is obtained from the temperature dependence of
$\rho$ in Eq.~(\ref{EFVRH}), $g(E)$ is the background DOS, i.e,
on which the Coulomb gap is superimposed,
and $C$ a dimensionless constant $\approx 6$. If we take $(\partial
\ln g(E)/\partial E)_{E_F} \sim 1/E_F$ (implying that $E_F$ may be
in the tail of the DOS) and again using $g_0$ to estimate $E_F$,
we find that the calculated $S^d$ are typically a factor of two
smaller than the values of $S$ at the observed minima. The
argument is not significantly changed if Mott VRH is
assumed.\cite{B&CVRH,Zvyagin} In this case $S^d \propto T^{1/3}$
but the  magnitudes calculated for $S^d$ are similar. The
references give the relevant details.

In the insulating regime, one expects a diverging localization or
correlation length as the critical point is approached. Given this,
one might also question whether the thermopower results are in the
linear region close to the critical point. However, on this side of
the transition we are never very close to the critical density. The
closest point is at about $n_0-n \approx 0.5\times 10^{14}$m$^{-2}$.
The correlation length estimated in Ref.~
\cite{correlation}
under these conditions is about 20$\,\mu$m. Assuming a magnitude of the disorder
potential of order 1\,K gives a threshold electric field for
non-linear effects of order 80\,mV/m whereas the thermoelectric
electric field is about 1\,mV/m over the range of 0.7-7\,K. Thus
non-linear effects do not appear to be a problem for the present data.

To put the present results in perspective,
as far as we are aware the only previous work which attempted
to follow $S^d$ into the region of 2D electron localization was that
of Burns and Chaikin\cite{B&C} on thin films of Pd and PdAu. They
found an upturn of $S^d$ in the strong localization region but no
divergence
at higher conductivities. The authors attributed their results to
the opening of a Mott-Hubbard gap. In 3D, Lauinger
and Baumann\cite{L&B} observed critical behaviour of $\sigma$ and
a divergence of $S^d$ for metallic AuSb films, but the magnitude of
the latter was 2 orders of magnitude smaller than seen here.
Other experiments on bulk SiP\cite{SiP} and NbSi\cite{NbSi}
saw no divergence of $S^d$ on the metallic side.

We close this section by making a few comments about $S$ at
higher $T$. $S^g$ appears to be dominant in this region
because the strong $T$ dependence of $S$ (roughly $T^2$ to
$T^3$) is inconsistent with any other mechanism. As we have
already
indicated, little is known about the behaviour of $S^g$ where
$n$ is low and $\rho$ is high,\cite{EP2DS,ql} and
these are just the conditions that pertain around $n_0$.
Thus, although $S^g$ should be present on the metallic side,
its  precise form is not known with any certainty.

When conduction is dominated by VRH, we expect $S^g =0$
because drag is based on the conservation of crystal momentum for
electron-phonon scattering which will not hold for transitions
between localized electron states.\cite{Zvyagin,P&F} Thus, $S^g$
on the insulator side should only exist when carrier excitation to
delocalized states occurs, though we emphasize that there are no
calculations appropriate to these conditions. Our $\rho$ data are
consistent with activated behaviour for $n$ just below $n_0$, but
deeper into the insulating state VRH becomes dominant and so we
would expect to see a strong diminution of $S^g$ in this region.
Figure~6 shows experimental data on $S$ as a function of $n$ at
fixed temperatures of 2\,K, 3\,K and 4\,K. We see that $S$ rises
as $n$ decreases but crosses $n_0$ smoothly, i.e., we no longer
see divergent behaviour of $S$ at $n_0$ as we did for $S^d$. Indeed
there is no feature that indicates  that anything unusual occurs
at $n_0$. This behaviour is consistent with activated conduction
below $n_0$. However, we continue to see a rise in $S$, and
presumably $S^g$, to the lowest densities, which is not expected
from our argument above. This can only be understood if
significant activated conduction is also present at all densities.
In this light, the fact that our $\rho$ data at lowest $n$ in
Fig.~2, (and the data of others\cite{VRHresistivity}) appear to
follow the Efros-Shklovskii VRH model so well over a limited range
of temperatures appears to be somewhat coincidental.

\section{Summary and Conclusions}
Generally speaking, the main focus of attention with regard to the
MIT in 2D systems has been whether the metallic behaviour is the
result of a transition to a new state induced by the strong
disorder and possibly electron-electron interactions, or whether
it is the result of a conventional physical mechanism that has yet
to be unambiguously identified.

On the metallic side, our analysis has mainly concentrated on data
at low $T$, essentially below the region of rapid $T$ dependence
of $\rho$. Our main result is that the critical behaviour of
$\sigma_0$ and $S^d$ of Si-MOSFETs in this region are surprisingly
consistent with equations that are formally equivalent to those
describing a 3D Anderson MIT. On the insulating side, the
behaviour of $\rho$ and $S^d$ over a wider range of $T$ are also
consistent with such a scenario. However, not all features that we
see can be understood in this way. In particular, although the
Anderson model exhibits scaling behaviour in both
$\sigma (T)$ and $S^d(T)$,\cite{Villagonzalo2}  it does not appear
to be able to explain the large increase of $\rho$ with $T$ in the
metallic region observed in Si-MOSFETs.

An Anderson transition can arise purely from disorder in 3D but
general scaling results predict that the equivalent transition
should not occur in 2D. Still, given that many features of the
data do mimic an Anderson transition, a key question in the
present work would then be whether a `mobility edge' actually
exists, in particular at low or zero temperature. If we look for
more conventional explanations of the data, a point to bear in
mind is that $k_Fl \sim 0.3 $  at $n_0$, where $l$ is the electron
mean free path estimated from the conductivity, which implies very
strong disorder. Under these conditions it seems plausible that a
sufficiently rapid drop in $\rho$ with decreasing $n$ could be
mistaken for a mobility edge, but it might actually be caused  by
a smooth, though very rapid, transition from  weak to strong
(exponential) localization of the carriers in keeping with the
scaling model. If this is the explanation, then it remains to be
shown how this rapid drop can mimic a critical behaviour over a
relatively wide range of $n$.

In our analysis, the apparent divergence of $S^d$ as $n_0$ is approached
from above is basically a
reflection of the rapid drop in $\sigma_0$, a result that might remain
valid even if no Anderson transition occurred.
An unambiguous indicator of a mobility edge would be saturation of $S^d$
as a function of temperature. Our data near $n_0$ show no  visible
indication of this, but the presence of phonon drag is a complicating
factor here and might mask any saturation. Clearly it would be an advantage
to suppress phonon drag and reveal $S^d$ over a wider range of $T$, but
this is difficult to do in 2D. In principle, it could be done by the use
of very thin substrates.
We do not see any features in the overall behaviour of $S$ which appear
to correlate with the strong $T$ dependence of $\rho$ in the metallic
region.

Finally, an important point that must  be stressed is that to reliably
identify any observed critical behaviour with a MIT requires data in
the zero $T$ limit. Although our analysis of $\sigma_0$ and $S^d$
is based on an extrapolation to zero $T$, the actual data extend
only to 0.3K. If the strong drop of $\rho$ is caused by conventional
physics, then we are examining the low temperature limit of this
mechanism,
but not necessarily the true low temperature limit of the system.
This is true of practically all the experimental data published so far.

In spite of these reservations, it is clear that the observed
critical behaviour in both $\sigma_0$ and $S^d$ indicates an
unknown but interesting physics of strongly disordered and
interacting systems. The results open up a new window on these
systems that further constrains any theoretical model proposed  to
explain the MIT, whether such a transition be a quantum property
or simply the result of a conventional physical mechanism.

\section{Acknowledgements}
We acknowledge the support of the NSERC of Canada, and from INTAS,
 NSF, RFBR, Programs \lq Physics of nanostructures',
\lq Statistical physics', \lq Integration' and `the State support
of the leading scientific schools'. We are also grateful for the
very useful comments by Rudolf R\"{o}mer and Cristine
Villagonzalo at Technische Universit\"{a}t, Chemnitz.

\newpage

\begin{figure}
\label{fig1} \caption{Examples of resistivity $\rho$ at various
densities (in units of $10^{15}$m$^{-2}$) as a function of
temperature $T$ for Sample 1 in the insulating region. The data
are plotted in the form $\log \rho$ as a function of $T^{-1/2}$ to
show that Eq.~(\ref{EFVRH}) (Efros-Shklovskii VRH) gives a
good representation at low $n$ (0.52 and 0.61 on the figure). At
higher $n$ (0.70, 0.79 and 0.88) the data are well
described by Eq.~(\ref{activated}) below 1\,K,
 corresponding to simple activated
behaviour. }
\end{figure}

\begin{figure}
\label{fig2}
\caption{Resistivity $\rho$ at various fixed densities $n$ (in units of
$10^{15}$m$^{-2}$) as a function of temperature $T$ for Sample 2 in the
metallic region. The solid lines are fits to Eq.~(\ref{metalrho}) using
only the points represented by open symbols. The closed symbols for the
data near $n_0$ are points not used in the fitting (see text). The
short vertical lines give the value of $T_0$ for each density.
The data at $n = 0.96$ are nominally at $n_0$.}
\end{figure}

\begin{figure}
\label{fig3}
\caption{The main panel shows the density
dependence of the conductivity $\sigma_0$ in the $T\rightarrow 0$
limit of Sample 2. The inset shows the same data for Sample 3.
In both cases circles correspond to data obtained using
Eq.~(\ref{metalrho}).
The lowest points designated by triangles are simply the measured
$\sigma$(T=0.3K). The dashed vertical lines are
the values of $n_0$ used in obtaining the fitted curves. In the
case of Sample 3, two fitted curves are shown (but are almost
indistiguinshable over the data range) corresponding to the two
values of $n_0$.}
\end{figure}

\begin{figure}
\label{fig4}
\caption{The thermopower $S$ for Sample 1 at various fixed
electron densities $n$ (in units of $10^{15}$\,m$^{-2}$)
as a function of temperature.}
\end{figure}

\begin{figure}
\label{fig5}
\caption{
Density dependence of  $\alpha = S^d/T$ for Sample 1.
The closed symbols are obtained using Eq.~(\ref{Sd+Sg})
and the open symbols are simply the measured $S/T$ at
at 0.3\,K (the values of $\alpha$ from the two methods
at $n > 3\times 10^{15}$\,m$^{-2}$ are indistinguishable).
The points at $n =0.97\times 10^{15}$\,m$^{-2}$
are just below $n_0$  but $k_BT$ broadening
should make these indistinguishable from $n_0$.
The line is the best fit to Eq.~(\ref{scalesd})
for $n < 4\times 10^{15}$\,m$^{-2}$.}
\end{figure}

\begin{figure}
\label{fig6}
\caption{ Thermopower at fixed temperatures of 2.0\,K, 3.0\,K and
4.0\,K as a function of density.}
\end{figure}

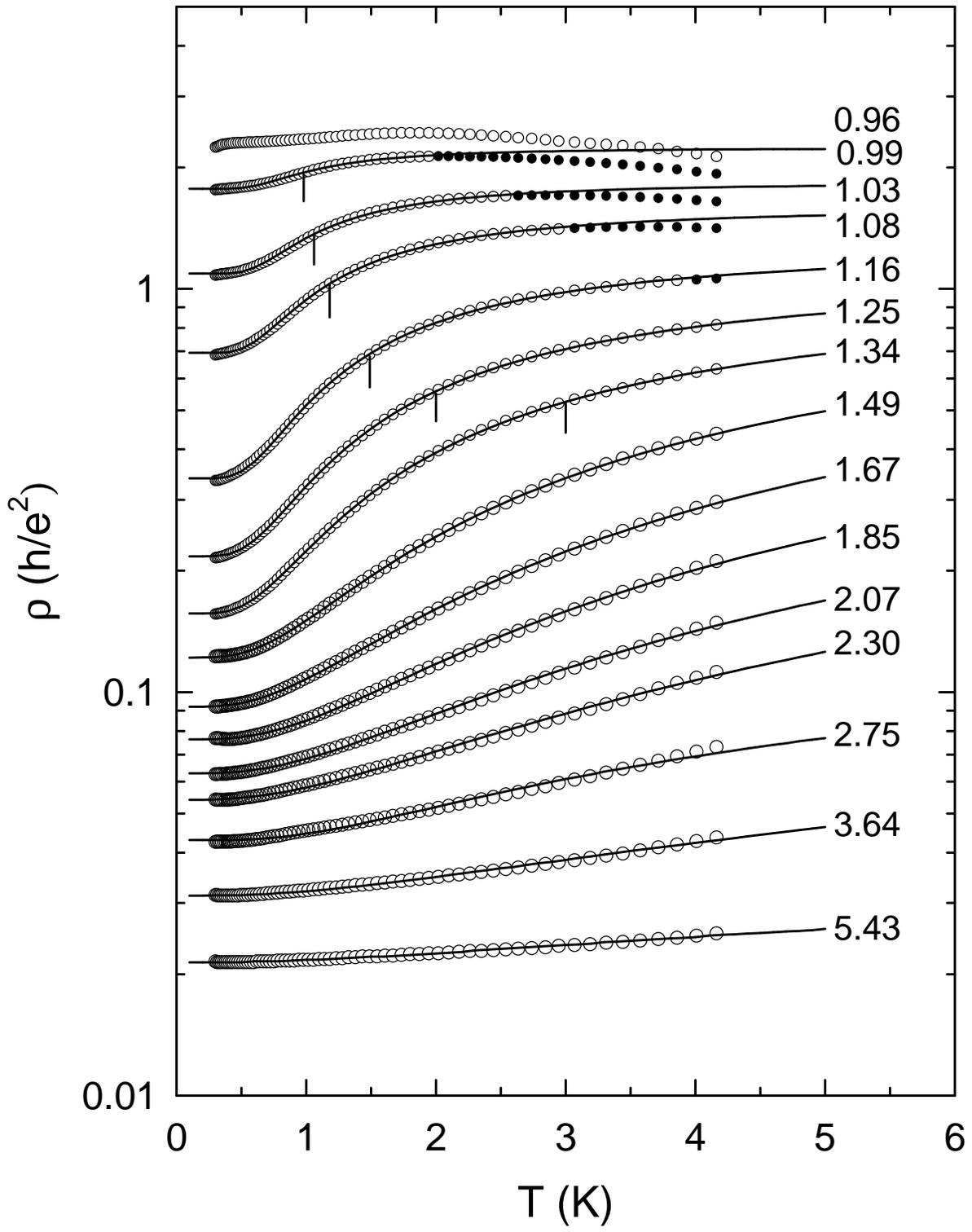
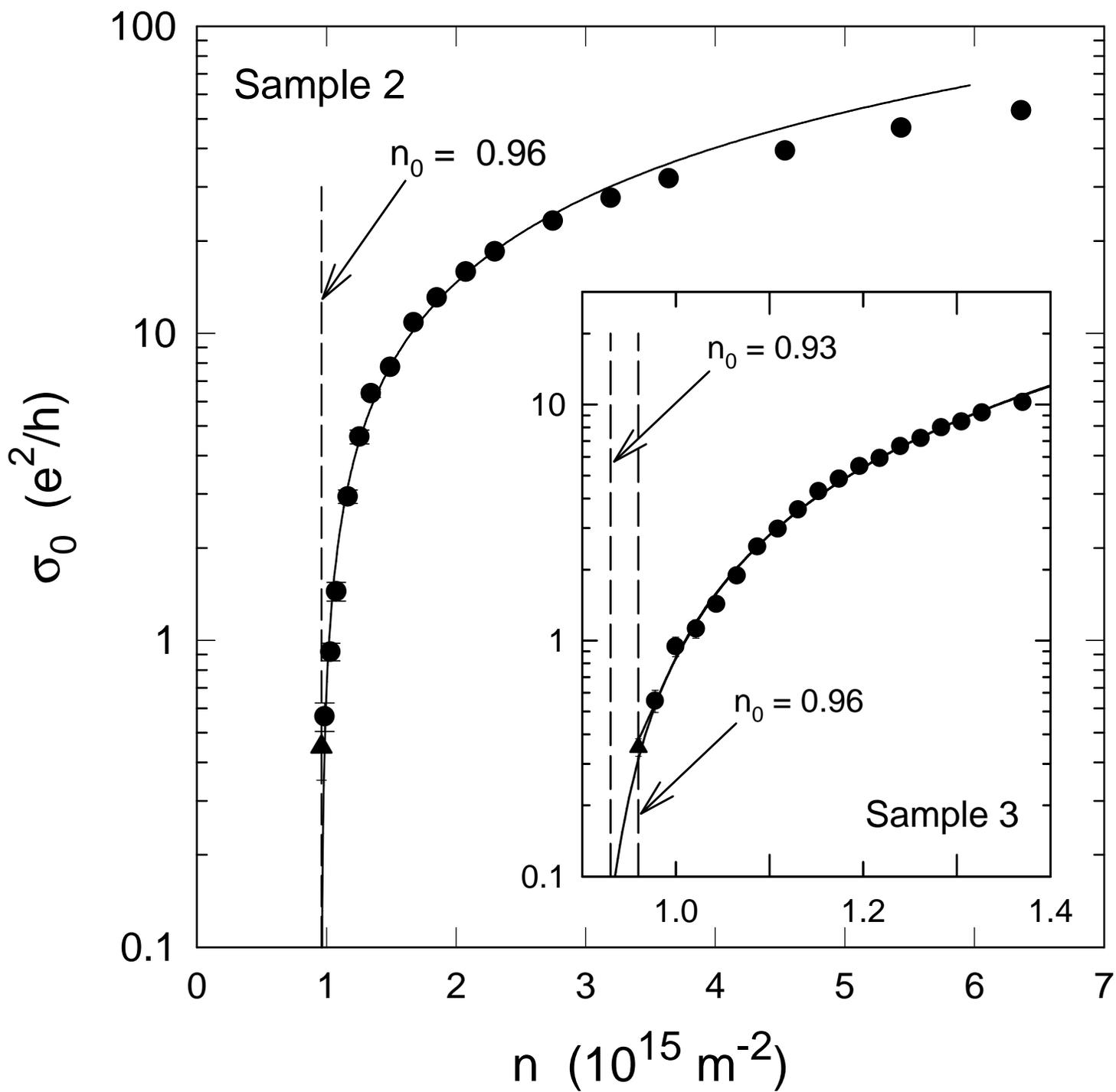
\end{document}